\documentclass[journal=cgdefu,manuscript=article]{achemso}
\usepackage{setspace,tabularx,caption,booktabs}
\usepackage{geometry, tabularx}
\usepackage{graphicx}
\usepackage{subcaption}
\usepackage{epstopdf}
\usepackage{pslatex, float}
\usepackage{gensymb}
\usepackage{xspace}
\usepackage{pifont}
\usepackage{multicol}
\usepackage{textcomp}
\usepackage{float}
\usepackage{natbib}
\usepackage[T1]{fontenc} 
\usepackage[usenames, dvipsnames]{color}

\newcommand{\cso}{CuSeO$_3$\xspace}
\newcommand{\coso}{Cu$_2$OSeO$_3$\xspace}
\newcommand{\cosoo}{Cu$_4$O(SeO$_3$)$_3$\xspace}

\author{Jessica R. Panella}
\author{Benjamin A. Trump}
\affiliation[ChemJHU]
{Department of Chemistry, Johns Hopkins University, Baltimore, MD 21218}
\author{Guy G. Marcus}
\affiliation[IQMPandA]
{Institute for Quantum Matter and the Department of Physics and Astronomy, Johns Hopkins University, Baltimore, MD 21218}
\author{Tyrel M. McQueen}
\affiliation[ChemJHU]
{Department of Chemistry, Johns Hopkins University, Baltimore, MD 21218}
\alsoaffiliation[IQMPandA]
{Institute for Quantum Matter and the Department of Physics and Astronomy, Johns Hopkins University, Baltimore, MD 21218}
\alsoaffiliation[MSE]
{Department of Materials Science and Engineering, Johns Hopkins University, Baltimore, MD 21218}
\email{mcqueen@jhu.edu}

\title{Seeded Chemical Vapor Transport Growth of \coso}

\begin{document}

\captionsetup{labelfont=bf,font=small,labelsep=period}

\maketitle

\begin{abstract}

We present an optimized seeded chemical vapor transport method for the growth of \coso that allows for chemical control in a system with many stable phases to selectively produce large phase pure single crystals.  This method is shown to consistently produce single crystals in the range of 120 to 180 mg.  A Wulff construction model of a representative crystal shows that the minimum energy surface is \{1~1~0\}, followed by \{1~0~0\}.  Analysis of the lowest index planes revealed that cleavage of Se-O bonds has a large energy cost, leading to an overall high surface energy.  The seeded chemical vapor transport demonstrated here shows promise for large single crystal growth of other functional materials such as Weyl semimetals, frustrated magnets, and superconductors.

\end{abstract}

\section{Introduction}

Investigation of magnetic materials is frequently limited by an inability to obtain sufficiently large, phase-pure single crystals.  This is especially true for materials containing anisotropic magnetism, geometric magnetic frustration, or ferrimagnetism.  A unique example of this is \coso, which has been studied heavily in recent years \cite{belesi_ferrimagnetism_2010, bos_magnetoelectric_2008, seki_observation_2012, janson_quantum_2014, portnichenko_magnon_2016,laurita_low_2017}.  Already a rare example of an insulating, noncentrosymmetric ferrimagnet \cite{bos_magnetoelectric_2008}, the production of single crystals was essential for the discovery of a magnetic skyrmion lattice \cite{seki_observation_2012,adams_long-wavelength_2012}.

The cubic \coso crystallizes in a distorted pyrochlore structure with Cu$_4$O and SeO$_3$ units.  Three Cu$^{2+}$ in each tetrahedron are ferromagnetically aligned and the remaining Cu$^{2+}$ is anti-parallel to the others, forming a net spin-1 entity \cite{bos_magnetoelectric_2008, romhanyi_entangled_2014}.  This ordering is the result of both symmetric and antisymmetric magnetic exchange interactions.  While the ground state magnetic ordering is ferrimagnetic, there is a region of parameter space for which helical spin textures emerge due to the Dzyaloshinskii-Moriya interaction leading to a magnetic skyrmion lattice \cite{skyrme_unified_1962,muhlbauer_skyrmion_2009}.  \coso is the only known insulating skyrmion material, which makes it a promising candidate for the realization of voltage-driven skyrmion-based devices and for use as a substrate material in thin film work.

\coso crystals sufficient for these experiments have previously been grown by chemical vapor transport (CVT), which uses a vapor phase transport agent to facilitate gas phase diffusion of one or more of the species involved in the growth across a controlled temperature gradient \cite{belesi_ferrimagnetism_2010, bos_magnetoelectric_2008,miller_magnetodielectric_2010}.  Typical transport agents include H$_2$, HCl, Cl$_2$, I$_2$, and TeCl$_4$ which react with one or more of the species involved with the growth to produce high vapor pressure species with high mobility \cite{mercier_recent_1982,binnewies_chemical_2013}.  Nucleation in CVT typically occurs along the wall of the reaction vessel, and can result in many small crystals if many nucleation sites are present.  Adding a seed crystal to the cool end of the gradient promotes nucleation on the seed, encouraging the growth of one large crystal.  There are few examples of seeded CVT in the literature, with the main ones being growth of transparent ZnO and  IR-sensitive Hg$_{0.8}$Cd$_{0.2}$Te \cite{skupinski_seeded_2010, wiedemeier_bulk_1991}.

Recent reports have used source temperatures of 610\degree C and 607\degree C and deposition temperatures of 500\degree C and 527\degree C, using HCl gas as the transport agent \cite{belesi_ferrimagnetism_2010,miller_magnetodielectric_2010, dyadkin_chirality_2014}.  The seeded CVT method reported herein improves the phase selectivity over previously reported methods via carefully chosen source and deposition temperatures, and the addition of a seed crystal dramatically increases the average size of the grown crystals.  The result is the ability to consistently produce large (\texttildelow 150 mg), high purity, extremely well-faceted crystals.  The size and quality of crystals grown by seeded CVT makes them well-suited for use as substrates due to the well-defined facets, and aids in the ability to align them for measurements such as terahertz spectroscopy, inelastic neutron scattering, and thermal transport \cite{laurita_low_2017, prasai_ballistic_2017-1}.

\section{Methods}
\subsection{Materials}
CuO (Alfa Aesar, 99.995\%) and NH$_4$Cl (Sigma-Aldrich, ACS Grade) were used as received.  SeO$_2$ (Alfa Aesar, 99.99\%) was purified by sublimation under flowing oxygen at 325 \degree C per Ref. \cite{waitkins_selenium_1945} to avoid off-stoichiometry and dangerously high pressures.  The oxygen flow was directed through a drying tube filled with Drierite then bubbled through concentrated nitric acid (BDH, ACS Grade) before passing over the as-received, pink SeO$_2$.  The resulting snow-white SeO$_2$ needle-like crystals were transferred to an argon-filled glovebox.  Care should be taken when handling SeO$_2$ due to its extreme toxicity.

\subsection{Synthesis of \coso Powder}
Polycrystalline \coso was synthesized by direct solid state reaction of the oxides, in a 2:1 molar ratio of CuO to SeO$_2$, following Bos \textit{et al.} \cite{bos_magnetoelectric_2008}.  The oxide precursors were ground thoroughly in an argon-filled glovebox and 5-20 g were sealed in an evacuated fused silica ampoule.  The reaction mixture was ramped to 600 \degree C at a rate of 100 \degree C/hr, held at temperature for 12 hrs, then the reaction tube was quenched in water.  The resulting product was ground in air and reheated until the product was a uniform olive green color upon grinding, usually three times.

\subsection{Seed Crystal Growth}  Single crystals were grown by the chemical vapor transport method using NH$_4$Cl as the transport agent \cite{belesi_ferrimagnetism_2010,portnichenko_magnon_2016}. A one gram charge of \coso powder was sealed in an evacuated fused silica ampoule along with 0.4 mg/cm$^3$ NH$_4$Cl, which was added in an argon-filled glovebox.  The tube was positioned between the first two zones of a three-zone tube furnace.  The source and deposition zones were ramped at 100 \degree C/hr to 640 \degree C and 530 \degree C, respectively.  The temperature accuracy of each zone was verified using an external, NIST calibration-traceable thermocouple.  This temperature gradient was maintained for 6 weeks.  The source end was cooled to room temperature approximately 7 hrs before the deposition end to allow inversion of the temperature gradient and force vapor phases to condense opposite the crystals.  Crystals were rinsed with Type I, deionized H$_2$O and ethanol once the growth finished to remove any remaining NH$_4$Cl.

\subsection{Seeded CVT Growth}
A crystal from an unseeded growth, weighing 10 mg $\leq m \leq$ 20 mg, was chosen for use as a seed crystal for a seeded growth.  Any obvious twinned or conjoined portion was cleaved off prior to use as a seed.  \coso powder and NH$_4$Cl were added to one end of a fused silica tube as before, taking care to keep the powder off of the tube walls.  The tube was necked in the middle, the seed crystal was added, and the tube was sealed under vacuum.  The same heating cycle was used for the seeded growth.  The tube set up and temperature gradient are shown in Fig. \ref{fig:crystal}(a).  The results of roughly 36 unseeded and 38 seeded growths are reported here.

\subsection{Characterization}  Back-reflection X-ray Laue diffraction was used to identify crystal faces and crystalline quality.  Laue images were taken using a Multiwire Laboratories MWL 110 real-time back reflection Laue camera with a tungsten source operating sub-K$_{\alpha}$ at 10 kV and 10 mA. Powder X-ray diffraction (XRD) of ground single crystals was used to confirm phase purity using a Bruker D8 Focus operating with a Cu K$_{\alpha}$ (K$_{\alpha1}$ = 1.5406 \AA) radiation with a LynxEye position sensitive detector.  Additionally, magnetization was also used to confirm phase purity and reproducibility of physical properties.  Magnetization data was collected on a Quantum Design Physical Properties Measurement System using the ACMS dc option.  Temperature-dependent magnetization was measured with an applied field of $\mu_0$H = 18 Oe, and field-field dependent magnetization was measured at T = 2 K and T = 60 K over the field range $\mu_0$H = 0 to 70000 Oe.

Structure images were generated using VESTA \cite{momma_vesta_2011}, and Wulff constructions were generated using WinXMorph \cite{kaminsky_winxmorph:_2005}. QLaue was used to simulate the Laue diffraction pattern \cite{wilkins_qlaue_nodate}.

\section{Results and Discussion}

\begin{figure}[h!]
	\includegraphics[width=3.33in]{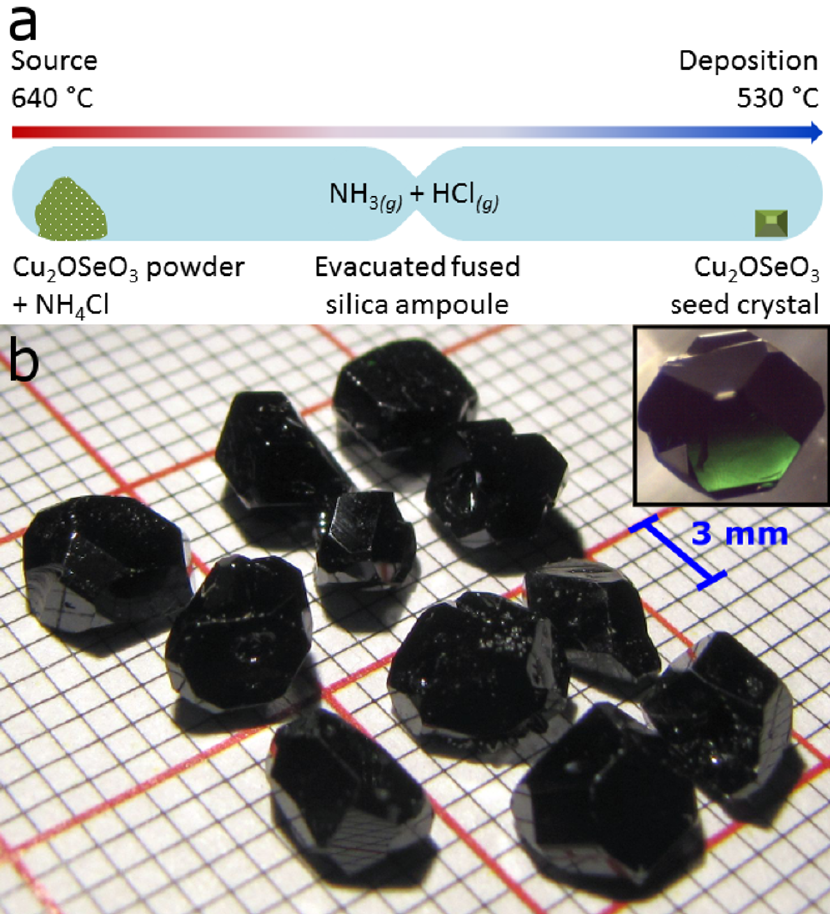}
	\caption{Seeded CVT Diagram and Resultant Crystals. (a) A diagram of the seeded CVT growth set-up.  (b) A collection of large \coso crystals grown by the seeded CVT method, displaying smooth, mirror-like facets.  The inset shows a small crystal set on a reflective sheet to highlight the color of the material.}
	\label{fig:crystal}
\end{figure}

\begin{figure}[h!]
	\includegraphics[width=3.33in]{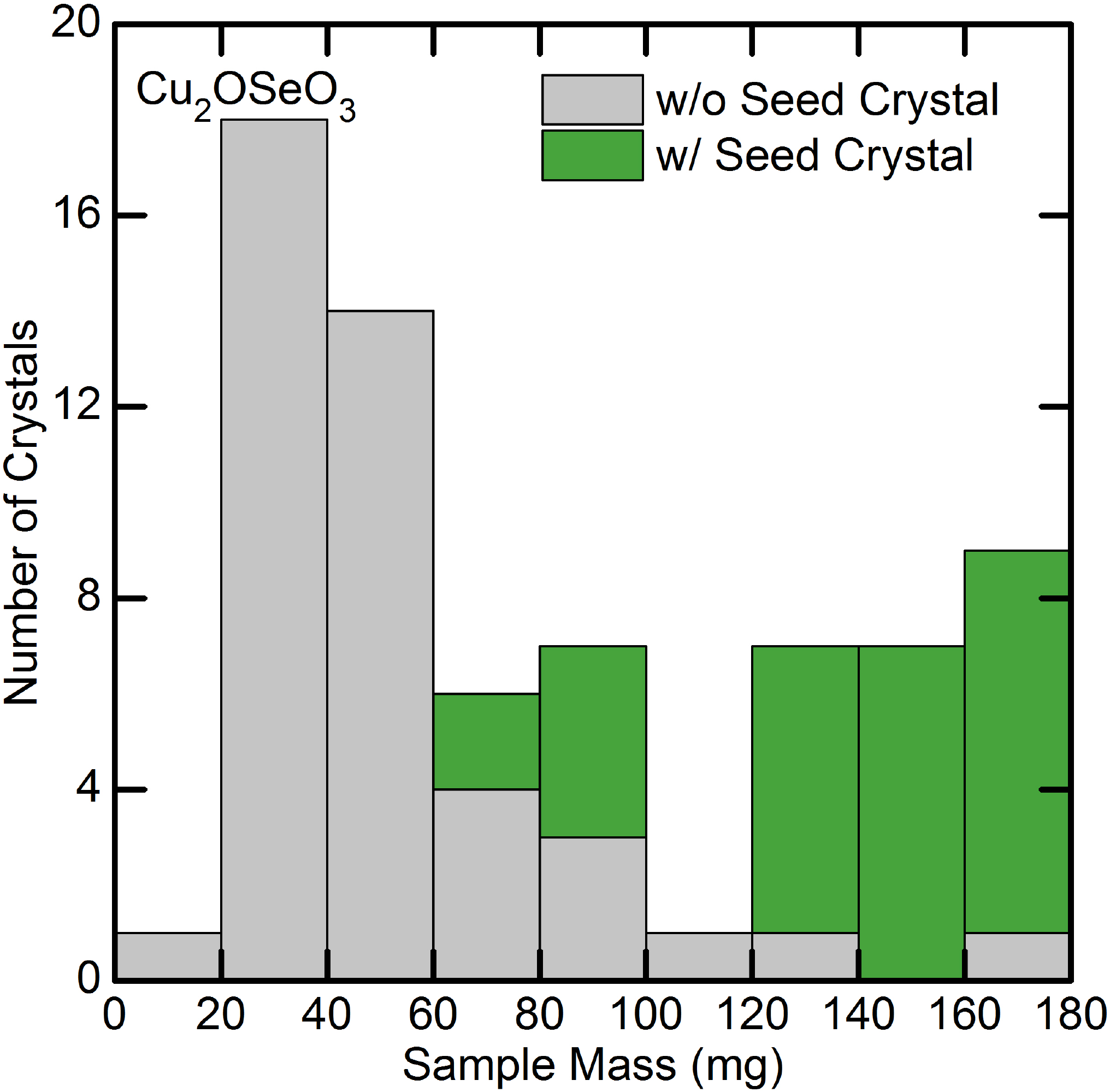}
	\caption{Histogram of Crystal Sizes.  A histogram showing the number of crystals having a given mass for about 36 unseeded and 38 seeded growths, with the bimodal distribution demonstrating the advantage of the seeded growth (green) over unseeded growth (gray) for obtaining large single crystals.}
	\label{fig:hist}
\end{figure}

A schematic diagram of the seeded CVT set up is shown in Fig. \ref{fig:crystal}(a), with the charge and transport agent shown on the source end and a single seed crystal on the deposition end.  The large, glossy, dark green crystals grown by seeded CVT, are shown in Fig. \ref{fig:crystal}(b) with the inset showing a small crystal set on a reflective sheet to highlight the color of the material.  Each growth produces roughly 100-150 mg of crystals (10-15\% yield), irrespective of the mass of the charge, either as one large (\texttildelow4x4x4mm) crystal or several smaller crystals.  Using seed crystals larger than \texttildelow20 mg did not additionally benefit the growth, while significantly smaller seeds could not initiate nucleation, likely because they were consumed by the transport agent during furnace warming. The masses of crystals grown by a similar amount of unseeded and seeded growths are summarized in the histogram in Fig. \ref{fig:hist}.  The histogram clearly shows a bimodal distribution in crystal sizes, with seeded growths (green) consistently yielding much larger crystals.  Unseeded growths (grey) produced many more crystals, as the overall yield by mass is the same despite much smaller sizes.  Crystals weighing less than 10 mg were excluded for clarity.

The CuO-SeO$_2$ system has several species that form under very similar thermodynamic conditions and differ in the ratio of CuO:SeO$_2$, namely \cso (1:1), \cosoo (4/3:1), and \coso (2:1) \cite{effenberger_kristallstrukturen_1986,meunier_constantes_1976}.  That they differ primarily in the ratio of two species makes phase selection based on stoichiometry difficult, as direct control of the ratio of the vapor phase species is nearly impossible.  Careful tuning of the reaction temperature provides conditions more favorable for the formation of \coso over related species, as detailed below.

The transport agent, NH$_4$Cl$_{(s)}$, sublimes at around $T$ = 340\degree C, yielding NH$_{3(g)}$ and HCl$_{(g)}$. At $T$ > 581.5\degree C, well below the source temperature (640\degree C), \coso decomposes into CuO$_{(s)}$ and SeO$_{2(g)}$ \cite{papankova_relationship_1989,fokina_thermodynamics_2014}.  Given the high vapor pressure of SeO$_2$ at elevated temperatures, we propose that SeO$_2$ self-transports, while CuO$_{(s)}$ reacts with HCl$_{(g)}$ to form volatile Cu-containing species that transport\cite{waitkins_selenium_1945}.  Nucleation at the deposition end continually removes vapor phase species, a process which is in dynamic equilibrium with the vaporization at the source end.  Our results are explained if the high source zone temperature increases the rate of reaction between CuO$_{(s)}$ and HCl$_{(g)}$ to achieve a higher ratio of CuO:SeO$_2$, closer to the ideal 2:1 ratio, which helps prevent formation of \cso.  The deposition zone temperature, 530\degree C, was chosen to be just higher than the decomposition temperature of the most commonly observed contaminant, \cosoo, which decomposes at $T$ $\geq$ 527.3\degree C \cite{fokina_thermodynamics_2014}.  This helps prevent concurrent growth of \cosoo and produces high purity unseeded and seeded crystals alike.  This is in contrast to previous growths of \coso, which used deposition temperatures below the decomposition of \cosoo and significantly lower source temperatures for an overall smaller temperature gradient \cite{belesi_ferrimagnetism_2010,miller_magnetodielectric_2010, dyadkin_chirality_2014}.

Furthermore, the use of a seed crystal helps prevent unwanted excess nucleation along the tube wall by locally depleting the concentration of gas-phase species as they deposit on the seed crystal.  The seed crystal also presents a more favorable substrate for deposition than silica.  The large temperature gradient (110\degree C) creates turbulence within the tube, effectively limiting the ability of small, undesired nuclei from growing to a stable size.  Additionally, since the seed already has the appropriate 2:1 ratio of CuO:SeO$_2$, further growth will maintain the stoichiometry, producing one large \coso crystal.

\begin{figure}[h!]
	\includegraphics[width=3.33in]{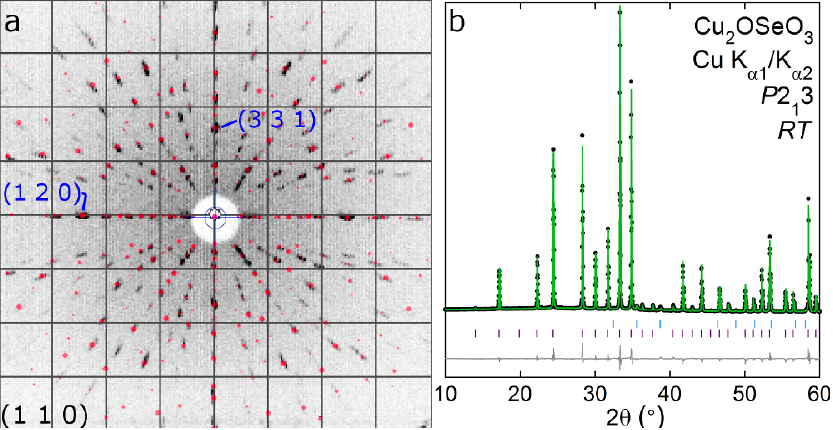}
	\caption{Single Crystal Laue XRD Pattern.  (a)  An experimental diffraction image taken looking at the (1~1~0) plane with a calculated pattern overlaid.  (b)  A powder XRD pattern demonstrating phase purity.  The data is shown in black, the calculated fit in green, and the difference is shown below in gray.  The dark violet tick marks correspond to allowed \coso peak positions, and the light blue ticks to CuO.}
	\label{fig:laue}
\end{figure}

The crystallinity of seeded CVT grown samples is demonstrated via the Laue XRD pattern shown in Fig. \ref{fig:laue}, oriented in the [1~1~0] direction.  Diffraction spots close to the center are exceptionally sharp, exemplifying the excellent crystallinity of these crystals.  Streaking seen on spots farther from the center is due to experimental broadening caused by the broad spectrum x-ray source of the MWL 110.  The experimental pattern shows no measurable mosaicity or twinning.  The calculated pattern, shown in Fig. \ref{fig:laue}(b), closely matches the experimental data.  To confirm that crystals did not contain core impurities, several were independently ground for powder XRD, and no other ternary Cu-Se-O phases were present within the detection limit of the diffractometer, as shown in Fig \ref{fig:laue}(c).  Rietveld refinement of a calculated fit (green) to the experimental data (black) using only \coso and CuO captures all of the visible peaks with an goodness of fit of 3.442.  The small amount of CuO present in the powder XRD pattern is likely unreacted powder from the surface of the crystal from condensation of residual gas phase Cu species.  Additionally, magnetization data (not shown) as a function of temperature and field, both above and below T$_N$, was used to confirm purity, consistently yielding data identical to those in Refs. \cite{bos_magnetoelectric_2008,seki_observation_2012,adams_long-wavelength_2012}.  Further, single crystal inelastic neutron scattering data supports bulk purity of the crystals \cite{marcus_neutrons_2016}.

\begin{figure}[h!]
	\includegraphics[width=3.33in]{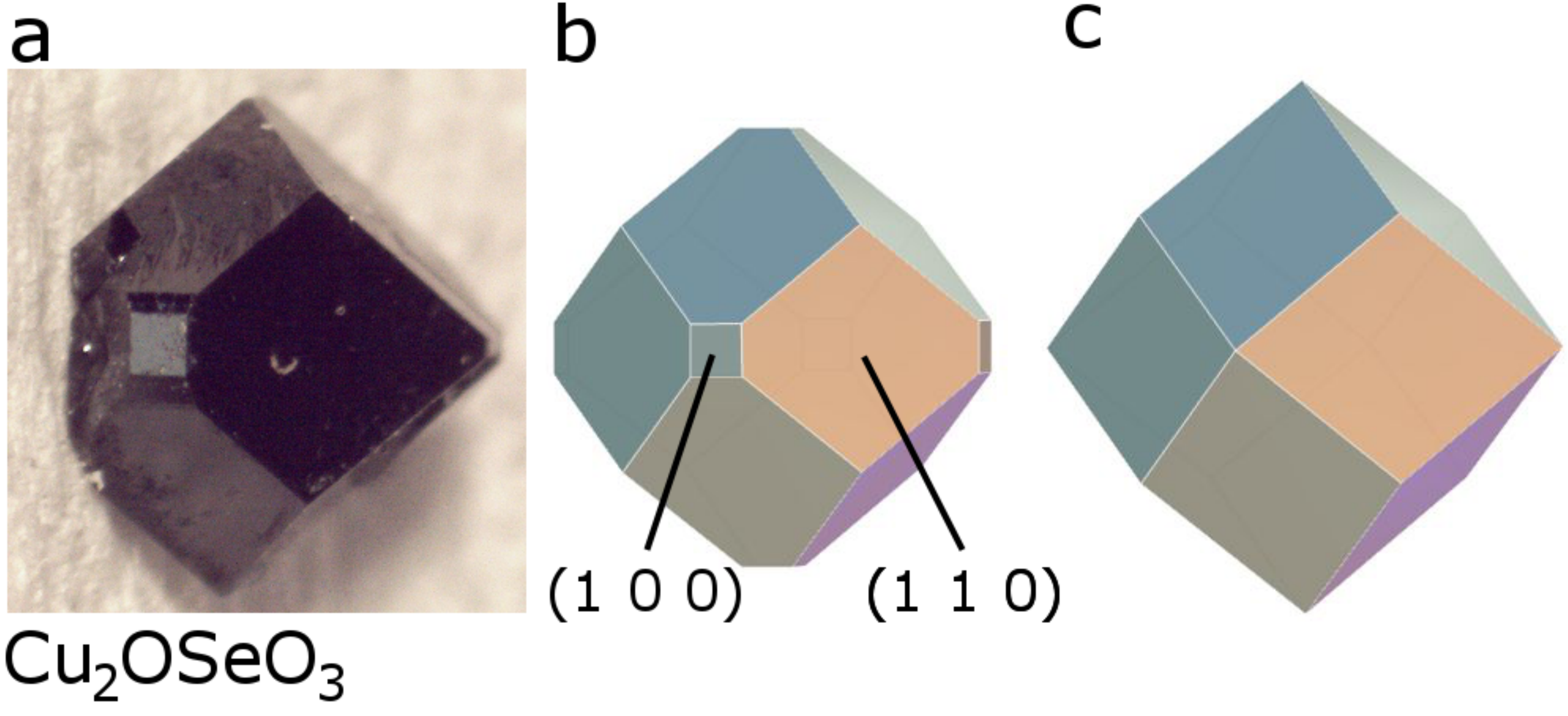}
	\caption{Crystal and Matching Wulff Construction (a) A representative crystal very near its equilibrium structure. (b) A Wulff construction matching the structure of the crystal. (c) Proposed crystal morphology in the absence of the influence of the tube wall.}
	\label{fig:wulff}
\end{figure}

The quality of the crystals is exemplified by their well-defined, dodecahedral forms, as expected in cubic crystal systems.  The exact shape of any crystal is controlled by the relative surface energies of the lattice planes, with the lowest surface energy plane defining the largest surface area face of the crystal.  Restated, the ratio of the surface areas of two or more crystal facets is inversely proportional to the ratio of surface energies for the corresponding lattice planes.  This principle is used in generating a Wulff construction, which can give a ratio of relative energies, though not an absolute energy \cite{wulff_zur_2015}.  In the case of \coso, the crystal morphology seems to be impacted by the crystals' adherence to the wall of reaction tube.  Growth from the tube wall could result in additional strain on the crystal during growth, changing the energy balance between the crystal planes enough to affect the global morphology of the crystal, as strong interactions with growth environment is known to affect morphology \cite{liu_morphology_1994,barmparis_nanoparticle_2015}.  Alternatively, interactions with the tube wall may change the growth kinetics, resulting in different crystal shapes depending on which plane adheres to the tube wall and how close the crystal is to reaching equilibrium.  

As an example of this effect, a well-formed seed crystal is shown in Fig. \ref{fig:wulff}(a) alongside a corresponding Wulff construction in Fig. \ref{fig:wulff}(b) that is defined by (1~1~0)- and (1~0~0)-type planes, and a similar construction that is defined by only (1~1~0)-type planes in Fig. \ref{fig:wulff}(c).  The crystal is almost an ideal rhombic dodecahedron, defined by primarily (1~1~0)-type faces, with one square (1~0~0)-type face truncating a single vertex.  The vertex that is occupied by the (1~0~0)-type face sits very close to the edge that adhered to the tube wall, while the other vertices that are situated farther away from the tube reach sharp points.  Assuming that the crystal is in its equilibrium shape, then the lowest surface energy plane for the bulk, free-standing crystal, in the absence of the tube wall, is the (1~1~0), and all other planes are significantly higher in energy, which would render a form as shown in Fig. \ref{fig:wulff}(c).  However, for the portion of the crystal that is strongly interacting with the fused silica tube, the surface energy of (1~0~0)-type planes is much closer to the surface energy of (1~1~0)-type planes, resulting in the growth of a (1~0~0)-defined facet.  If the effect of the tube wall is kinetic rather than thermodynamic, then Fig. \ref{fig:wulff}(c) is the equilibrium shape and further growth would result in the disappearance of the (1~0~0) face.

Using the Wulff construction in Fig. \ref{fig:wulff}(b), the ratio of surface energies in the strongly interacting region of the crystal is 1.26:1 (1~0~0):(1~1~0).  This ratio qualitatively matches what is observed by visual inspection of the crystals, which are all affected by fused silica growth tube.  Most of the crystals have large (1~1~0)-type faces and smaller (1~0~0)-type faces.  Occasionally, other low-index planes, such as (1~1~1), will be displayed, which is probably determined by the orientation of the crystals during the nucleation process.

\begin{figure}[h!]
	\includegraphics[width=6in]{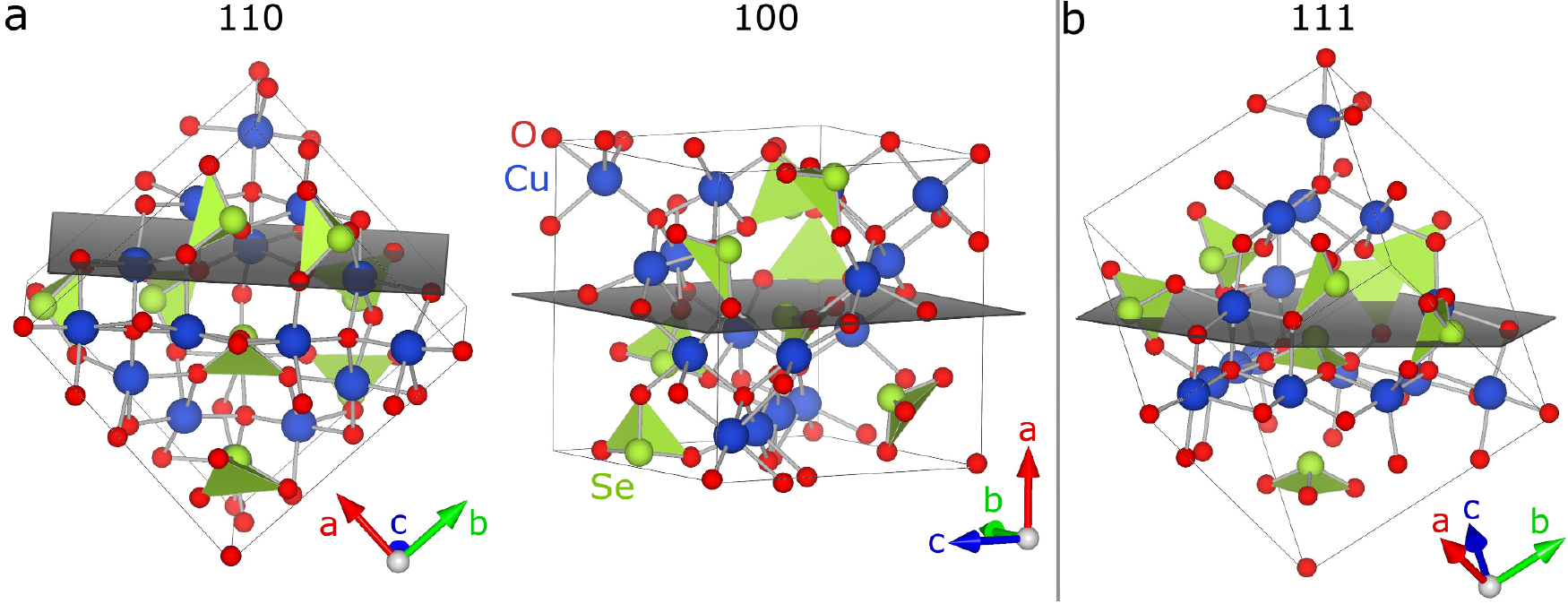}
	\caption{Structure with Low Index Planes.  (a) Both the (1~1~0)- and (1~0~0)-type planes are shown at a position within the cell where all SeO$_3$ polyhedra are either fully above or fully below the plane. (b) The (1~1~1) is placed to minimize the number of SeO$_3$ bonds broken, but it still bisects multiple SeO$_3$ polyhedra.}
	\label{fig:planes}
\end{figure}

In the absence of strong environmental effects, the lattice planes have characteristic surface energies that are affected by a number of system-dependent chemical and interface factors, such as lattice relaxation, bond termination, and surface reconstruction processes.  While all systems experience some degree of relaxation or reconstruction at interfaces, an examination of the crystal structure of \coso yields insight into the relationship between bond energy and surface energy: namely that the two predicted lowest energy planes, the (1~1~0) and the (1~0~0), can cleave the lattice without breaking any Se-O bonds.  The next lowest index plane, the (1~1~1), must either break Se-O bonds or cleave the densely-packed layers of Cu-O polyhedra.

While any dangling bond is reactive and high energy, the bond dissociation energy of an Se-O bond is 430(6) kJ/mol as compared to 287(11) kJ/mol for a Cu-O bond, implying that a surface containing a broken Se-O bond would be much higher in energy than a surface containing only broken Cu-O bonds \cite{luo_comprehensive_2007}.  Fig. \ref{fig:planes} shows the structure of \coso with each of the three lowest index planes placed to minimize the effect on the rigid SeO$_3$ units.  Both the (1~1~0) and (1~0~0) planes shown in Fig. \ref{fig:planes}(a) are placed a narrow region in the structure that does not contain any SeO$_3$, demonstrated by the face that the green SeO$_3$ polyhedra are all fully above or fully below the black crystal planes.  The (1~1~1) plane in Fig. \ref{fig:planes}(b) intersects multiple SeO$_3$ polyhedra, even when placed to minimize Se-O bond cleavage.

Based on observations of the shape of the many crystals produced by both seeded and unseeded CVT, the (1~1~0) and (1~0~0) are both low energy surfaces, indicating that bond cleavage is likely a strong factor in determining the surface energies in this system, with the Se-O bond cleavage being highly unfavorable.  Given that neither the (1~1~0) or (1~0~0) planes necessarily break Se-O bonds, other factors, such as Cu-O bond cleavage and surface reconstruction, likely determine which has the least surface energy.  An understanding of surface behavior in the \coso system is important to further studies in thin film and interface physics.

\section{Conclusion}
The seeded CVT method reported here optimizes crystal growth conditions for species selectivity and for producing large \coso single crystals based on an understanding of both the vapor phase chemistry and equilibrium thermodynamics.  Careful selection of source and deposition zone temperatures provides conditions that are unfavorable for competing Cu-Se-O species, while the use of a seed crystals inhibits excessive nucleation and reinforces the 2:1 CuO:SeO$_2$ ratio to encourage growth of large, single-phase \coso crystals.  Seeded CVT produces crystals that frequently weigh in excess of 100 mg and have exceptionally well-defined faces and clean edges.  Crystals grown by seeded CVT are of sufficient quality for a plethora of measurements and techniques, from inelastic neutron scattering to use as substrate material for thin film work.  Furthermore, the recognizable dodecahedral morphology of the crystals gives insight into the relative surface energies of the underlying crystal structure and extrinsic factors affecting those energies.  The techniques and chemical insight reported herein are applicable to numerous electronically and magnetically interesting systems currently grown by traditional, unseeded CVT, such as Weyl semimetals WTe$_2$, MoTe$_2$, ZrTe$_5$, TaP, NbP, superconductor FeSe, and spin-liquid candidate FeSc$_2$S$_4$ \cite{ali_large_2014,qi_superconductivity_2016,sapkota_single_2016,tsurkan_structure_2017,hara_structural_2010,lv_microstructure_2017}.

\section{Acknowledgments}
This research was supported by the Research Corporation for
Science Advancement (Cottrell Scholar Award) and NSF, Division of Materials Research, Solid State Chemistry, CAREER Grant under Award DMR-1253562.  GGM acknowledges generous support from the NSF-GRFP, Grant No. DGE-1232825.

\bibliography{Bib6}

\end{document}